\documentclass[twocolumn,aps,pra,groupedaddress,showpacs,floatfix]{revtex4}
\usepackage{amssymb,amsmath}
\usepackage{graphicx}

\begin{document}

\title{Efficient and perfect state transfer in quantum chains}

\author{Daniel Burgarth$^{1}$, Vittorio Giovannetti$^{2}$, and Sougato
Bose$^{1}$}

\affiliation{$^{1}$ Department of Physics \& Astronomy, University College London,
Gower St., London WC1E 6BT, UK. \\
 $^{2}$ NEST-INFM \& Scuola Normale Superiore, piazza dei Cavalieri
7, I-56126 Pisa, Italy.}

\begin{abstract}
We present a communication protocol for chains of permanently coupled
qubits which achieves perfect quantum state transfer and which is
efficient with respect to the number chains employed in the scheme.
The system consists of $M$ uncoupled identical quantum chains. Local control (gates, measurements)
is only allowed at the sending/receiving end of the chains.
Under a quite general
hypothesis on the interaction Hamiltonian of the qubits a theorem is proved which shows that the receiver 
 is able to asymptotically recover the messages by 
repetitive monitoring of his qubits.

\end{abstract}


\maketitle

\section{Introduction}

Permanently coupled quantum chains have recently been proposed 
as prototypes of reliable quantum communication lines~\cite{BOSE,LLOYD}.
The main drawback of these schemes is related with the fact that 
even in the absence of external noise the fidelity of the
transmission is in general not optimal~\cite{BOSE,GIOVA,key-11,key-19,key-38,PLENIO04}.
This is due to the dispersion which affects the
propagation of local excitations~\cite{OSBORNE}. 
One way to overcome this is to engineer specific coupling Hamiltonians~\cite{key-11,key-33,key-34,key-17,key-12}.
However, the more a scheme relies on particular properties of the
Hamiltonian, the more it will be affected by imperfections in its
implementation \cite{key-19}. A more general approach was taken in
\cite{key-3} where a specific encoding using time-dependent couplings
at the sending and receiving end of the chain achieved high fidelity transfer.
Perfect transfer (i.e. unitary fidelity) for a whole class of 
unmodulated quantum chains was finally achieved in Ref.~\cite{BURGARTH} 
by employing a parallel channel encoding where the sender of the message is 
able to transmit one qubit of information 
by operating on the first spins of two non interacting copies of the chain. 
In quantum information theory 
the ratio $R$ between the  number qubits that can be
transferred  with unitary fidelity and 
the number of channel copies used in the protocol is an important efficiency
parameter~\cite{SHOR,GIOVA}.
Therefore one question that naturally arises is whether or not there is any special
meaning in the 1/2 value of $R$ achieved in the protocol of Ref.~\cite{BURGARTH}.
More specifically we pose the following question:
can we use almost \emph{any} quantum chain for perfect and efficient (i.e. $R=1$)
quantum communication? In this article, we give a sufficient and easily
attainable condition for achieving this goal.

The paper is organized as follows: the model and the notation are
introduced in Sec.~\ref{s:sec1}. The efficiency and the fidelity of the protocol are discussed 
in Sec.~\ref{s:sec2} and in Sec.~\ref{s:sec3}, respectively. Finally in Sec.~\ref{s:sec4}  we  prove
a theorem which provides us with a sufficient condition for achieving efficient and perfect
state transfer in quantum chains.

\begin{figure}
\begin{center}\includegraphics[%
  width=0.90\columnwidth]{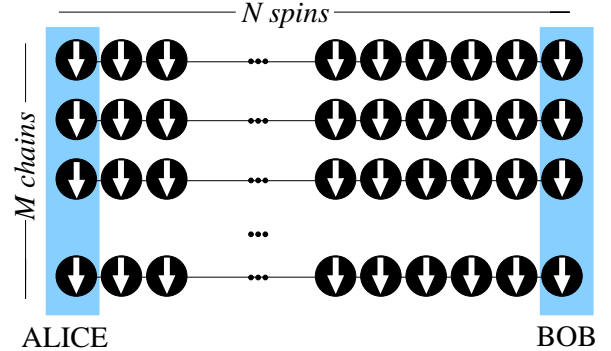}\end{center}

\caption{Schematic of the system: Alice and Bob operate $M$ chains, each
containing $N$ spins. The spins belonging to the same chain interact
through the Hamiltonian $H$ which accounts for the transmission of
the signal in the system. Spins of different chains do not interact.
Alice encodes the information in the first spins of the chains by
applying unitary transformations to her qubits. Bob recovers the
message in the last spins of the chains by performing joint measurements.}

\label{f:fig1}
\end{figure}

\section{The model}\label{s:sec1}

Consider a linear chain of
$N$ spins interacting through the Hamiltonian $H$. For $n=1,\cdots N$,
we define the single excitation vector 
\begin{eqnarray}
|\boldsymbol{n}\rangle\equiv|00\cdots010\cdots0\rangle\;,\label{excitation}\end{eqnarray}
 as the state of the chain in which the $n$-th spin is in the computational
base vector $|1\rangle$ and the remaining $N-1$ qubits are in the
state $|0\rangle$. Analogously we define $|{\boldsymbol{0}}\rangle\equiv|00\cdots0\rangle$
to be the state where all spins are in $|0\rangle$. We assume that
$|{\boldsymbol{0}}\rangle$ is an eigenvector of $H$ and that the
$N$-dimensional subspace generated by the states $|\boldsymbol{n}\rangle$
is invariant under the time evolution $u(t)\equiv e^{-iHt/\hbar}$,
i.e. \begin{eqnarray}
|\boldsymbol{n}\rangle\longrightarrow u(t)|\boldsymbol{n}\rangle=\sum_{n^{\prime}=1}^{N}f_{{n^{\prime}},{n}}(t)|{\boldsymbol{n^{\prime}}}\rangle\;,\label{OUTPUT}\end{eqnarray}
 where $f_{{n^{\prime}},{n}}(t)\equiv\langle{\boldsymbol{n^{\prime}}}|e^{-iHt/\hbar}|\boldsymbol{n}\rangle$
is the probability amplitude that the excitation $|\boldsymbol{n}\rangle$
moves to $|\boldsymbol{n^{\prime}}\rangle$ in the time interval $t$.
A sufficient criterion for Eq.~(\ref{OUTPUT})
is that $H$ commutes with the z component of the total spin. A typical example is provided by a linear array of spins with
Heisenberg interaction. 
In the original proposal of Ref.~\cite{BOSE} one assumes that initially
the chain is in $|\boldsymbol{0}\rangle$ and that at time $t=0$
a first party (Alice) encodes one qubit of logical information in
the first spin by preparing the chain in $|\Psi\rangle\equiv\alpha|\boldsymbol{0}\rangle+\beta|\boldsymbol{1}\rangle$
with $\alpha$ and $\beta$ complex. By reading out the state of the
$N$-th qubit at time $t$ a second party (Bob) will be able to
recover the information transmitted. 


Assume now that the two communicating parties operate on $M$ independent
(i.e. non interacting) copies of the chain~\cite{PARALLEL}.
The idea is to use these
copies to improve the overall fidelity of the communication. 
As in the original scheme~\cite{BOSE} we assume Alice and Bob to control
respectively the first and last qubit of each chain (see Fig.~\ref{f:fig1}).
By preparing any superposition of her spins Alice can in principle
transfer up to $M$ logical qubits. However, in order to improve the
communication fidelity the two parties will find it more convenient
to redundantly encode only a small number (say $Q(M)\leqslant M$)
of logical qubits in the $M$ spins. By adopting these strategies
Alice and Bob are effectively sacrificing the efficiency $R(M)=Q(M)/M$
of their communication line in order to increase its fidelity. This
is typical of any communication scheme and it is analogous to what
happens in quantum error correction theory, where a single logical
qubit is stored in many physical qubits. 
By focusing on those strategies that guarantee a
(possibly asymptotic in $M$) unitary fidelity in the transmission
of the $Q(M)$ encoded qubits, the efficiency $R(M)$ yields the capacity
of the channel~\cite{SHOR}.
In the case of the quantum chains~(\ref{OUTPUT}) it has been proved~\cite{BURGARTH} 
the existence of an encoding of efficiency $R=1/2$ which allows for unitary fidelity,
by showing 
that for $M=2$ it is possible to achieve perfect state transfer of
a single logical qubit by using just two copies of the original chain.
Here we will generalize such result by proving that given $M>2$ there exist an
optimal encoding-decoding strategy which asymptotically allows to
achieve perfect state transfer of $Q(M)$ qubits such that
\begin{eqnarray}
\lim_{M\rightarrow\infty}R(M)=1\;.\label{efficiency}\end{eqnarray}
In other words we show the possibility of achieving both perfect transfer 
\emph{and} optimal efficiency.

Our strategy requires Alice to prepare superpositions of the $M$
chains where $\sim M/2$ of them have a single excitation in the first
location while the remaining are in $|{\boldsymbol{0}}\rangle$. Since
in the limit $M>>1$ the number of qubit transmitted is $\log\binom{M}{M/2}\approx M$,
this architecture guarantees optimal efficiency~(\ref{efficiency}).
On the other hand, our protocol requires Bob to perform collective
measurements on his spins to determine if all the $\sim M/2$ excitations
Alice is transmitting arrived at his location. We will prove that
by repeating these detections many times, Bob is able to recover
the messages with asymptotically perfect fidelity.

\subsection{Notation}\label{s:sub1}

\begin{figure}
\begin{center}\includegraphics[%
  width=0.60\columnwidth]{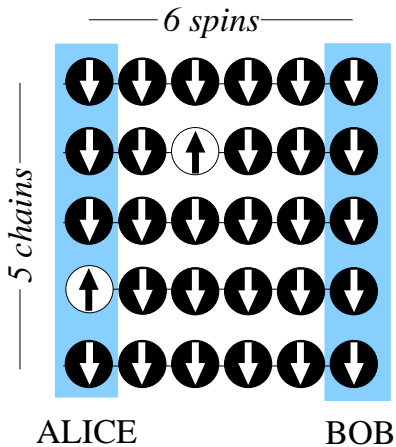}\end{center}

\caption{\label{cap:Example-of-our}Example of our notation for $M=5$ chains
of length $N=6$ with $K=2$ excitations. The state above, given by
$|\boldsymbol{0}\rangle_{1}\otimes|\boldsymbol{3}\rangle_{2}\otimes|\boldsymbol{0}\rangle_{3}\otimes|\boldsymbol{1}\rangle_{4}\otimes|\boldsymbol{0}\rangle_{5},$
has excitations in the chains $m_{1}=2$ and $m_{2}=4$ at the horizontal
position $n_{1}=3$ and $n_{2}=1$. It is in the Hilbert space $\mathcal{H}(S_{6})$
corresponding to the subset $S_{6}=\{2,4\}$ (assuming that the sets
$S_{\ell}$ are ordered in a canonical way, i.e. $S_{1}=\{1,2\},$ $S_{2}=\{1,3\}$
and so on) and will be written as $|(3,1);6\rangle\!\rangle.$ There
are $\binom{5}{2}=10$ different sets $S_{\ell}$ and the number of qubits
one can transfer using these states is $\log_{2}10\approx3.$ The
efficiency is thus given by $R\approx3/5$ which is already bigger
than in the original scheme \cite{BURGARTH}.}
\end{figure}

Before beginning the analysis let us introduce some notation. In order
to distinguish the $M$ different chains we introduce the label $m=1,\cdots,M$:
in this formalism $|\boldsymbol{n}\rangle_{m}$ represents the state~(\ref{excitation})
of $m$-th chain with a single excitation in the $n$-th spin. In
the following we will be interested in those configurations of the
whole system where $K$ chains posses a single excitation while the
remaining $M-K$ are in $|\boldsymbol{0}\rangle$, as in the case\begin{equation}
|\boldsymbol{1}\rangle_{1}\otimes|\boldsymbol{1}\rangle_{2}\cdots\otimes|\boldsymbol{1}\rangle_{K}\otimes|\boldsymbol{0}\rangle_{K+1}\cdots\otimes|\boldsymbol{0}\rangle_{M}\label{eq:example}\end{equation}
 where for instance the first $K$ chains have an excitation in the first chain
location. Another more general example is given in Fig. \ref{cap:Example-of-our}.
The complete characterization of these vectors is obtained by specifying
\emph{i)} \emph{which} chains possess a single excitation and \emph{ii)}
\emph{where} these excitations are located horizontally along the
chains. In answering to the point \emph{i)} we introduce the $K$-element
subsets $S_{\ell}$, composed by the labels of those chains that contain
an excitation. Each of these subsets $S_{\ell}$ corresponds to a subspace
of the Hilbert space $\mathcal{H}(S_{\ell})$ with a dimension $N^{K}.$
The total number of such subsets is equal to the binomial coefficient
$\binom{M}{K}$, which counts the number of possibilities in which
$K$ objects (excitations) can be distributed amongst $M$ parties
(parallel chains). In particular for any $\ell=1,\cdots,\binom{M}{K}$
the $\ell$-th subset $S_{\ell}$ will be specified by assigning its
$K$ elements, i.e. $S_{\ell}\equiv\{ m_{1}^{(\ell)},\cdots,m_{K}^{(\ell)}\}$
with $m_{j}^{(\ell)}\in\{1,\cdots,M\}$ and $m_{j}^{(\ell)}<m_{j+1}^{(\ell)}$
for all $j=1,\cdots,K$. To characterize the location of the excitations,
point \emph{ii)}, we will introduce instead the $K$-dimensional vectors
$\vec{n}\equiv(n_{1},\cdots,n_{K})$ where $n_{j}\in\{1,\cdots,N\}$.
We can then define \begin{eqnarray}
|\boldsymbol{\vec{n}};\ell\rangle\!\rangle\equiv\bigotimes_{j=1}^{K}|\boldsymbol{n_{j}}\rangle_{m_{j}^{(\ell)}}\;\bigotimes_{m^{\prime}\in{\overline{S}_{\ell}}}|\boldsymbol{0}\rangle_{m^{\prime}}\;,\label{nvec}\end{eqnarray}
 where $\overline{S}_{\ell}$ is the complementary of $S_{\ell}$
to the whole set of chains.%
 The state (\ref{nvec}) represents a configuration where the $j$-th
chain of the subset $S_{\ell}$ is in $|\boldsymbol{n_{j}}\rangle$
while the chains that do not belong to $S_{\ell}$ are in $|\boldsymbol{0}\rangle$ (see
Fig. \ref{cap:Example-of-our} for an explicit example). The kets
$|\boldsymbol{\vec{n}};\ell\rangle\!\rangle$ are a natural generalization
of the states 
$|\boldsymbol{n}\rangle_{1}\otimes|\boldsymbol{0}\rangle_{2}$
and $|\boldsymbol{0}\rangle_{1}\otimes|\boldsymbol{n}\rangle_{2}$
which were used for the {}``dual-rail encoding'' in \cite{BURGARTH}.
They are useful for our purposes because they are mutually orthogonal,
i.e. \begin{eqnarray}
\!\langle\!\langle\boldsymbol{\vec{n}};\ell|\boldsymbol{\vec{n}^{\prime}};\ell^{\prime}\rangle\!\rangle=\delta_{\ell\ell^{\prime}}\;\delta_{\vec{n}\vec{n}^{\prime}}\;,\label{ortho}\end{eqnarray}
and their time evolution under the Hamiltonian does not depend on $\ell$ (cf. Eq.
(\ref{time})). Among the vectors~(\ref{nvec}) those where all the
$K$ excitations are located at the beginning of the $S_{\ell}$ chains
play an important role in our analysis. Here $\vec{n}=\vec{1}\equiv(1,\cdots,1)$
and we can write \begin{eqnarray}
|\boldsymbol{\vec{1}};\ell\rangle\!\rangle\equiv\bigotimes_{m\in S_{\ell}}|\boldsymbol{1}\rangle_{m}\;\bigotimes_{m^{\prime}\in{\overline{S}_{\ell}}}|\boldsymbol{0}\rangle_{m^{\prime}}\;.\label{in}\end{eqnarray}
 According to Eq.~(\ref{ortho}), for $\ell=1,\cdots,\binom{M}{K}$
these states form orthonormal set of $\binom{M}{K}$ elements. Analogously
by choosing $\vec{n}=\vec{N}\equiv(N,\cdots,N)$ we obtain the orthonormal
set of $\binom{M}{K}$ vectors \begin{equation}
|\boldsymbol{\vec{N}};\ell\rangle\!\rangle\equiv\bigotimes_{m\in S_{\ell}}|\boldsymbol{N}\rangle_{m}\;\bigotimes_{m^{\prime}\in{\overline{S}_{\ell}}}|\boldsymbol{0}\rangle_{m^{\prime}},\end{equation}
where all the $K$ excitations are located at the end of the chains.

\section{Efficient encoding}\label{s:sec2}

If all the $M$ chains of the system are originally in $|\boldsymbol{0}\rangle$,
the vectors~(\ref{in}) can be prepared by Alice by locally operating
on her spins. Moreover since these vectors span a $\binom{M}{K}$
dimensional subspace, Alice can encode in the chain $Q(M,K)=\log_{2}\binom{M}{K}$
qubits of logical information by preparing the superpositions, \begin{eqnarray}
|\Phi\rangle\!\rangle=\sum_{\ell}A_{\ell}\;|\boldsymbol{\vec{1}};\ell\rangle\!\rangle\;,\label{logicstate}\end{eqnarray}
 with $A_{\ell}$ complex coefficients. The efficiency of such encoding
is hence $R(M,K)=\frac{\log_{2}\binom{M}{K}}{M}$ which maximized
with respect to $K$ gives, \begin{eqnarray}
R(M) & = & \frac{1}{M}\left\{ \begin{array}{ll}
{\log_{2}\binom{M}{M/2}} & \;\mbox{for $M$ even}\\
{\log_{2}\binom{M}{(M-1)/2}} & \;\mbox{for $M$ odd}\;.\end{array}\right.\label{efficiencymax}\end{eqnarray}
 The Stirling approximation can then be used to prove that this encoding
is asymptotically efficient~(\ref{efficiency}) in the limit of large
$M$, e.g. \begin{eqnarray}
\log_{2}\binom{M}{M/2} & \approx & \log_{2}\frac{M^{M}}{(M/2)^{M}} =M.\end{eqnarray}
Note that already for $M=5$ the encoding is more efficient (cf. Fig.
\ref{cap:Example-of-our}) than in the {}``dual-rail encoding''
given in \cite{BURGARTH}. In the remaining of the paper we show that
the encoding~(\ref{logicstate}) provides  perfect state transfer
by allowing Bob to perform joint measurements at its end of the chains.

\section{Perfect transfer}\label{s:sec3}

Since the $M$ chains do not interact with each other and possess
the same free Hamiltonian $H$ (this assumption can be relaxed, see
\cite{key-32}), the unitary evolution of the whole system is described
by $U(t)\equiv\otimes_{m}u_{m}(t)$, with $u_{m}(t)$ being the operator~(\ref{OUTPUT})
acting on the $m$-th chain. The time evolved of the input $|\boldsymbol{\vec{1}};\ell\rangle\!\rangle$
of Eq.~(\ref{in}) is thus equal to \begin{eqnarray}
 &  & U(t)|\boldsymbol{\vec{1}};\ell\rangle\!\rangle=\sum_{\vec{n}}F[\vec{n},\vec{1};t]\;|\boldsymbol{\vec{n}};\ell\rangle\!\rangle\;,\label{time}\end{eqnarray}
 where the sum is performed for all $n_{j}=1,\cdots,N$ and \begin{eqnarray}
F[{\vec{n},\vec{n^{\prime}}};t]\equiv f_{n_{1},n_{1}^{\prime}}(t)\cdots f_{n_{K},n_{K}^{\prime}}(t)\;,\label{capitolF}\end{eqnarray}
 is a quantity which does \emph{not} depend on $\ell$. In Eq.~(\ref{time})
the term ${\vec{n}}=\vec{N}$ corresponds to having all the $K$ excitations
in the last locations of the chains. We can thus write \begin{eqnarray}
U(t)|\boldsymbol{\vec{1}};\ell\rangle\!\rangle=\gamma_{1}(t)|\boldsymbol{\vec{N}};\ell\rangle\!\rangle+\sqrt{1-|\gamma_{1}(t)|^{2}}\;|\boldsymbol{\xi}(t);\ell\rangle\!\rangle\;,\label{time1}\end{eqnarray}
 where \begin{eqnarray}
\gamma_{1}(t) & \equiv & \langle\!\langle\boldsymbol{\vec{N}};\ell|U(t)|\boldsymbol{\vec{1}};\ell\rangle\!\rangle=F[\vec{N},\vec{1};t]\label{gamma1}\end{eqnarray}
 is the probability amplitude that all the $K$ excitation of $|\boldsymbol{\vec{1}};\ell\rangle\!\rangle$
arrive at the end of the chains, and \begin{eqnarray}
|\boldsymbol{\xi}(t);\ell\rangle\!\rangle\equiv\sum_{\vec{n}\neq\vec{N}}F_{1}[\vec{n},\vec{1};t]\;|\boldsymbol{\vec{n}};\ell\rangle\!\rangle\;,\label{error}\end{eqnarray}
 with \begin{equation}
F_{1}[\vec{n},\vec{1};t]\equiv\frac{F[\vec{n},\vec{1};t]}{\sqrt{1-|\gamma_{1}(t)|^{2}}},\label{eq:13b}\end{equation}
is a superposition of terms where the number of excitations arrived
to the end of the communication line is strictly less then $K$. It
is worth noticing that Eq.~(\ref{ortho}) yields the following relations,
\begin{eqnarray}
\langle\!\langle\boldsymbol{\vec{N}};\ell|\boldsymbol{\xi}(t);\ell^{\prime}\rangle\!\rangle=0,\quad\!\langle\!\langle\boldsymbol{\xi}(t);\ell|\boldsymbol{\xi}(t);{\ell^{\prime}}\rangle\!\rangle=\delta_{\ell\ell^{\prime}}\;,\label{ortho1}\end{eqnarray}
which shows that $\left\{ ||\boldsymbol{\xi}(t);\ell\rangle\!\rangle\right\} $
is an orthonormal set of vectors which spans a subspace orthogonal
to the states $|\boldsymbol{\vec{N}};\ell\rangle\!\rangle.$ The time
evolution of the input state~(\ref{logicstate}) follows by linearity
from Eq.~(\ref{time1}), i.e. \begin{eqnarray}
|\Phi(t)\rangle\!\rangle=\gamma_{1}(t)\;|\Psi\rangle\!\rangle+\sqrt{1-|\gamma_{1}(t)|^{2}}\;|\overline{\Psi}(t)\rangle\!\rangle\;,\label{logicstateout}\end{eqnarray}
 with 
\begin{eqnarray}
|\overline{\Psi}(t)\rangle\!\rangle&\equiv&\sum_{\ell}A_{\ell}\;|\boldsymbol{\xi}(t);\ell\rangle\!\rangle\;,
\nonumber \\
|\Psi\rangle\!\rangle & \equiv & \sum_{\ell}A_{\ell}\;|\boldsymbol{\vec{N}};\ell\rangle\!\rangle\;.\label{ok}\end{eqnarray}
 The vectors $|\Psi\rangle\!\rangle$ and $|\overline{\Psi}(t)\rangle\!\rangle$
are unitary transformations of the input message~(\ref{logicstate})
where the orthonormal set $\{|\boldsymbol{\vec{1}};\ell\rangle\!\rangle\}$
has been rotated into $\{|\boldsymbol{\vec{N}};\ell\rangle\!\rangle\}$
and $\{|\boldsymbol{\xi}(t);\ell\rangle\!\rangle\}$ respectively.
Moreover $|\Psi\rangle\!\rangle$ is the configuration we need to have for
perfect state transfer at the end of the chain. In fact it is obtained
from the input message~(\ref{logicstate}) by replacing the components
$|\boldsymbol{1}\rangle$ (excitation in the first spin) with $|\boldsymbol{N}\rangle$
(excitation in the last spin). From Eq.~(\ref{ortho1}) we know that
$|\Psi\rangle\!\rangle$ and $|\overline{\Psi}(t)\rangle\!\rangle$
are orthogonal. This property helps Bob to recover the message $|\Psi\rangle\!\rangle$
from $|\Phi(t)\rangle\!\rangle$: he needs only to perform a collective
measurement on the $M$ spins he is controlling to establish if there
are $K$ or less excitations in those locations. The above
is clearly a projective measure that can be performed without destroying
the quantum coherence associated with the coefficients $A_{\ell}$.
Formally this can described by introducing the observable \begin{eqnarray}
\Theta\equiv\openone-\sum_{\ell}|\boldsymbol{\vec{N}};\ell\rangle\!\rangle\langle\!\langle\boldsymbol{\vec{N}};\ell|\;.\label{observable}\end{eqnarray}
 A single measure of $\Theta$ on $|\Phi(t_{1})\rangle\!\rangle$
yields the outcome $0$ with probability $p_{1}\equiv|\gamma_{1}(t_{1})|^{2}$,
and the outcome $+1$ with probability $1-p_{1}$. In the first case
the system will be projected in $|\Psi\rangle\!\rangle$ and Bob will
get the message. In the second case instead the state of the system
will become $|\overline{\Psi}(t_{1})\rangle\!\rangle$. Already at
this stage the two communicating parties have a success probability
equal to $p_{1}$. Moreover, as in \cite{BURGARTH}, the channels
have been transformed into a quantum erasure channel \cite{key-1}
where the receiver knows if the transfer was successful. 

Consider
now what happens when Bob fails to get the right answer from the measure.
The state on which the chains is projected is explicitly given by
\begin{eqnarray}
|\overline{\Psi}(t_{1})\rangle\!\rangle & = & \sum_{\vec{n}\neq\vec{N}}F_{1}[{\vec{n},\vec{1}};t_{1}]\sum_{\ell}A_{\ell}|\boldsymbol{\vec{n}};\ell\rangle\!\rangle\;.\label{explicit}\end{eqnarray}
 Let us now consider the evolution of
this state for another time interval $t_{2}$.  By repeating the
same analysis given above we obtain an expression similar to (\ref{logicstateout}),
i.e. \begin{eqnarray}
|\Phi(t_{2},t_{1})\rangle\!\rangle & = & \gamma_{2}\;|\Psi\rangle\!\rangle+\sqrt{1-|\gamma_{2}|^{2}}\;|\overline{\Psi}(t_{2},t_{1})\rangle\!\rangle\;,\label{logicstateout2}\end{eqnarray}
 where now the probability amplitude of getting all excitation in
the $N$-th locations is described by 
$$\gamma_{2}\equiv\sum_{\vec{n}\neq\vec{N}}F[{\vec{N},\vec{n}};t_{2}]\; F_{1}[{\vec{n},\vec{1}};t_{1}].$$
In this case $|\overline{\Psi}(t)\rangle\!\rangle$ is replaced by
\begin{eqnarray}
|\overline{\Psi}(t_{2},t_{1})\rangle\!\rangle & = & \sum_{\ell}A_{\ell}\;|\boldsymbol{\xi}(t_{2},t_{1});\ell\rangle\!\rangle\;,\label{ko2}\end{eqnarray}
 with 
$$|\boldsymbol{\xi}(t_{2},t_{1});\ell\rangle\!\rangle=\sum_{\vec{n}\neq\vec{N}}F_{2}[\vec{n},\vec{1};t_{2},t_{1}]|\boldsymbol{\vec{n}};\ell\rangle\!\rangle,$$
and $F_{2}$ defined as in Eq.~(\ref{effeq}) (see below). In other
words the state $|\Phi(t_{2},t_{1})\rangle\!\rangle$ can be obtained
from Eq.~(\ref{logicstateout}) by replacing $\gamma_{1}$ and $F_{1}$
with $\gamma_{2}$ and $F_{2}$. Bob can hence try to use the same
strategy he used at time $t_{1}$: i.e. he will check whether or not
his $M$ qubits contain $K$ excitations. With (conditional) probability
$p_{2}\equiv|\gamma_{2}|^{2}$ he will get a positive answer and his
quantum register will be projected in the state $|\Psi\rangle\!\rangle$
of Eq.~(\ref{ok}). Otherwise he will let the system evolve for another
time interval $t_{3}$ and repeat the protocol. Reiterating the above
analysis it is possible to give a recursive expression for the conditional
probability of success $p_{q}\equiv|\gamma_{q}|^{2}$ after $q-1$
successive unsuccessful steps. The quantity $\gamma_{q}$ is the analogous
of $\gamma_{2}$ and $\gamma_{1}$ of Eqs.~(\ref{gamma1}) and (\ref{logicstateout2}).
It is given by \begin{eqnarray}
\gamma_{q}\equiv\sum_{\vec{n}\neq\vec{N}}F[{\vec{N},\vec{n}};t_{q}]\; F_{q-1}[\vec{n},\vec{1},t_{q-1},\cdots,t_{1}]\;,\label{gammaq}\end{eqnarray}
 where \begin{eqnarray}
\lefteqn{F_{q-1}[\vec{n},\vec{1};t_{q-1},\cdots,t_{1}]}\label{effeq}\\
 & \equiv & \sum_{\vec{n}^{\prime}\neq\vec{N}}\frac{F[{\vec{N},\vec{n}^{\prime}};t_{q-1}]}{\sqrt{1-|\gamma_{q-1}|^{2}}}F_{q-2}[{\vec{n}^{\prime},\vec{1}};t_{q-2},\cdots,t_{1}]\nonumber \end{eqnarray}
and $F_{1}[\vec{n},\vec{1},t]$ is given by Eq. (\ref{eq:13b}). In
these equations $t_{q},\cdots,t_{1}$ are the \emph{times intervals}
that occurred between the various protocol steps. Analogously the
conditional probability of failure at the step $q$ is equal to $1-p_{q}$.
The probability of having $j-1$ failures and a success at the step
$j$-th can thus be expressed as \begin{eqnarray}
\pi(j) & = & p_{j}(1-p_{j-1})(1-p_{j-2})\cdots(1-p_{1})\;,\label{proba}\end{eqnarray}
 while the total probability of success after $q$ steps is obtained
by the sum of $\pi(j)$ for all $j=1,\cdots,q$, i.e. \begin{eqnarray}
P_{q} & = & \sum_{j=1}^{q}\pi(j)\;.\label{probtot}\end{eqnarray}
 Since $p_{j}\geqslant0$, Eq.~(\ref{probtot}) is a monotonic function
of $q$. As a matter of fact in the next section we prove that under a very general
hypothesis on the system Hamiltonian, the probability of success $P_{q}$
converges to $1$ in the limit of $q\rightarrow\infty$.
This means that by repeating many times
the collective measure described by $\Theta$ Bob is guaranteed to
get, sooner or later, the answer $0$ and hence the message Alice
sent to him. In other words our protocol allows perfect state transfer
in the limit of repetitive collective measures. Notice that the above
analysis applies for all classes of subsets $S_{\ell}$. The only
difference between different choices of $K$ is in the velocity of
the convergence of $P_{q}\rightarrow1$. In any case, by choosing
$K\sim M/2$ Alice and Bob can achieve perfect fidelity \emph{and}
optimal efficiency.

\section{Convergence theorem}\label{s:sec4}

Here we show that if there exists no eigenvector $|e_{m}\rangle$
of the quantum chain Hamiltonian $H$ which is orthogonal to $|\boldsymbol{N}\rangle$,
than there is a choice of the times intervals $t_{q},t_{q-1},\cdots,t_{1}$
such $P_{q}$ of Eq.~(\ref{probtot}) converges to $1$ in the limit
of $q\rightarrow\infty$. For the special case $M=2$ and $K=1$ this
was numerically shown in Ref.~\cite{BURGARTH}.

The state of the system at a time interval of $t_{q}$ after the $(q-1)$-th failure
can be expressed in compact form as follows \begin{eqnarray*}
|\Phi(t_{q},\cdots,t_{1})\rangle\!\rangle & = & \frac{U(t_{q})\Theta U(t_{q-1})\Theta\cdots U(t_{1})\Theta|\Phi\rangle\!\rangle}{\sqrt{(1-p_{q-1})\cdots(1-p_{1})}}\end{eqnarray*}
 with $U(t)$ the unitary time evolution generated by the system Hamiltonian,
and with $\Theta$ the projection defined in Eq.~(\ref{observable}).
One can verify for instance that for $q=2$, the above equation coincides
with Eq.~(\ref{logicstateout2}). {[}For $q=1$ this is just (\ref{logicstateout})
evaluated at time $t_{1}${]}. By definition the conditional probability
of success at step $q$-th is equal to 
$$p_{q}\equiv|\langle\!\langle\Psi|\Phi(t_{q},\cdots,t_{1})\rangle\!\rangle|^{2}.$$
Therefore, Eq.~(\ref{proba}) yields \begin{eqnarray}
\pi(q) & = & |\langle\!\langle\Psi|U(t_{q})\Theta U(t_{q-1})\Theta\cdots U(t_{1})\Theta|\Phi\rangle\!\rangle|^{2}\label{proba0}\\
 & = & |\langle\!\langle\boldsymbol{\vec{N}};\ell|U(t_{q})\Theta U(t_{q-1})\Theta\cdots U(t_{1})\Theta|\boldsymbol{\vec{1}};\ell\rangle\!\rangle|^{2}\;,\nonumber \end{eqnarray}
 where the second identity stems from the fact that, according to
Eqs.~(\ref{OUTPUT}) and (\ref{ortho}), $U(t)\Theta$ preserves
the orthogonality relation among states $|\boldsymbol{\vec{n}};\ell\rangle\!\rangle$
with distinct values of $\ell.$ Analogously to the cases of Eqs.~(\ref{capitolF})
and (\ref{gamma1}), the second identity of~(\ref{proba0}) establishes
that $\pi(q)$ can be computed by considering the transfer of the
input $|\boldsymbol{\vec{1}};\ell\rangle\!\rangle$ for \emph{arbitrary}
$\ell$. The expression (\ref{proba0}) can be further simplified
by noticing that for a given $\ell$ the chains of the subset $\overline{S}_{\ell}$
contribute with a unitary factor to $\pi(q)$ and can be thus neglected
(according to~(\ref{in}) they are prepared in $|\boldsymbol{0}\rangle$
and do not evolve under $U(t)\Theta$). Identify $|\boldsymbol{\vec{1}}\rangle\!\rangle_{\ell}$
and $|\boldsymbol{\vec{N}}\rangle\!\rangle_{\ell}$ with the components
of $|\boldsymbol{\vec{1}};\ell\rangle\!\rangle$ and $|\boldsymbol{\vec{N}};\ell\rangle\!\rangle$
relative to the chains belonging to the subset $S_{\ell}$. In this
notation we can rewrite Eq.~(\ref{proba0}) as \begin{eqnarray}
\pi(q) & = & |_{\ell}\!\langle\!\langle\boldsymbol{\vec{N}}|U_{\ell}(t_{q})\Theta_{\ell}\;\cdots U_{\ell}(t_{1})\Theta_{\ell}|\boldsymbol{\vec{1}}\rangle\!\rangle_{\ell}|^{2}\;,\label{proba1}\end{eqnarray}
 where $\Theta_{\ell}=\openone_{\ell}-|\boldsymbol{\vec{N}}\rangle\!\rangle_{\ell}\langle\!\langle\boldsymbol{\vec{N}}|$
and $U_{\ell}(t)$ is the unitary operator $\otimes_{m\in S_{\ell}}u_{m}(t)$
which describes the time evolution of the chains of $S_{\ell}$.

To prove that there exist a suitable choices of $t_{j}$ such that
the series~(\ref{probtot}) converges to $1$ it is sufficient to consider the case
$t_{j}=\tau>0$ for all $j=1,\cdots,q$: this is equivalent to selecting
decoding protocols with constant measuring intervals.

By introducing
the operator $T_{\ell}\equiv U_{\ell}(\tau)\Theta_{\ell}$, Eq.~(\ref{proba1})
becomes thus\begin{eqnarray}
 &  & \pi(q)=|_{\ell}\!\langle\!\langle\boldsymbol{\vec{N}}|\;(T_{\ell})^{q}|\boldsymbol{\vec{1}}\rangle\!\rangle_{\ell}|^{2}\label{proba2}\\
 &  & =_{\ell}\!\!\langle\!\langle\boldsymbol{\vec{1}}|(T_{\ell}^{\dag})^{q}|\boldsymbol{\vec{N}}\rangle\!\rangle_{\!\ell}\!\langle\!\langle\boldsymbol{\vec{N}}|\;(T_{\ell})^{q}|\boldsymbol{\vec{1}}\rangle\!\rangle_{\ell}=w(q)-w(q+1)\;, \nonumber \end{eqnarray}
 where \begin{equation}
w(j)\equiv_{\ell}\!\langle\!\langle\boldsymbol{\vec{1}}|(T_{\ell}^{\dag})^{j}\;(T_{\ell})^{j}|\boldsymbol{\vec{1}}\rangle\!\rangle_{\ell}=\Vert(T_{\ell})^{j}|
\boldsymbol{\vec{1}}\rangle\!\rangle_{\ell}\Vert^{2}\;, \end{equation}
is the norm of the vector $(T_{\ell})^{j}|\boldsymbol{\vec{1}}\rangle\!\rangle_{\ell}$.
Substituting Eq.~(\ref{proba2}) in Eq.~(\ref{probtot}) yields
\begin{eqnarray}
P_{q} & = & \sum_{j=1}^{q}\left[w(j)-w(j+1)\right]=1-w(q+1)\label{probtot1}\end{eqnarray}
 where the property $w(1)={_{\ell}\langle\!\langle}\boldsymbol{\vec{1}}|\Theta_{\ell}|\boldsymbol{\vec{1}}\rangle\!\rangle_{\ell}=1$
was employed. Proving the thesis is hence equivalent to prove that
for $q\rightarrow\infty$ the succession $w(q)$ nullifies. This last
relation can be studied using properties of power bounded matrices~\cite{INEQ}.
In fact, by introducing the norm of the operator $(T_{\ell})^{q}$
we have, \begin{eqnarray}
w(q)=\Vert(T_{\ell})^{q}|\boldsymbol{\vec{1}}\rangle\!\rangle_{\ell}\Vert^{2}\leqslant\Vert(T_{\ell})^{q}\Vert^{2}\leqslant c\left(\frac{1+\rho(T_{\ell})}{2}\right)^{2q}\label{thesis}\end{eqnarray}
 where $c$ is a positive constant which does not depend on $q$ \cite{key-36}
and where $\rho(T_{\ell})$ is the spectral radius of $T_{\ell}$,
i.e. the eigenvalue of $T_{\ell}$ with maximum absolute value (N.B.
even when $T_{\ell}$ is not diagonalizable this is a well defined
quantity). Equation~(\ref{thesis}) shows that $\rho(T_{\ell})<1$
is a sufficient condition for $w(q)\rightarrow0$. In our case we
note that, given any normalized eigenvector $|\lambda\rangle\!\rangle_{\ell}$
of $T_{\ell}$ with eigenvalue $\lambda$ we have \begin{eqnarray}
|\lambda|=\Vert T_{\ell}|\lambda\rangle\!\rangle_{\ell}\Vert=\Vert\Theta_{\ell}|\lambda\rangle\!\rangle_{\ell}\Vert\leqslant1\;,\label{thesis1}\end{eqnarray}
 where the inequality follows from the fact that $\Theta_{\ell}$
is a projector. Notice that in Eq.~(\ref{thesis1}) the identity
holds only if $|\lambda\rangle\!\rangle$ is also an eigenvector of
$\Theta_{\ell}$ with eigenvalue $+1$, i.e. only if $|\lambda\rangle\!\rangle_{\ell}$
is orthogonal to $|\boldsymbol{\vec{N}}\rangle\!\rangle_{\ell}$.
By definition $|\lambda\rangle\!\rangle_{\ell}$ is eigenvector $T_{\ell}=U_{\ell}(\tau)\Theta_{\ell}$:
therefore the only possibility to have the equality in Eq.~(\ref{thesis1})
is that \emph{i)} $|\lambda\rangle\!\rangle_{\ell}$ is an eigenvector
of $U_{\ell}(\tau)$ (i.e. an eigenvector of the Hamiltonian $H_{\ell}^{\mbox{\small{tot}}}$
of the chain subset $S_{\ell}$) and \emph{ii)} it is orthogonal to
$|\boldsymbol{\vec{N}}\rangle\!\rangle_{\ell}$. By negating the above
statement we get a sufficient condition for the thesis. Namely, if
all the eigenvectors $|\vec{E}\rangle\!\rangle_{\ell}$ of $H_{\ell}^{\mbox{\small{tot}}}$
are not orthogonal to $|\boldsymbol{\vec{N}}\rangle\!\rangle_{\ell}$
than the absolute values of the eigenvalues $\lambda$ of $T_{\ell}$
are strictly smaller than $1$ which implies $\rho(T_{\ell})<1$ and
hence the thesis. Since the $S_{\ell}$ channels are identical and
do not interact, the eigenvectors $|\vec{E}\rangle\!\rangle_{\ell}\equiv\bigotimes_{m\in S_{\ell}}|e_{m}\rangle_{m}$
are tensor product of eigenvectors $|e_{m}\rangle$ of the single
chain Hamiltonians $H$. Using the notation introduced in Eq.~(\ref{excitation})
 the sufficient condition becomes 
\begin{eqnarray}
_{\ell}\langle\!\langle\vec{E}|\boldsymbol{\vec{N}}\rangle\!\rangle_{\ell}
=\prod_{m\in S_{\ell}}{_{m}\!\langle \boldsymbol{N}}|e_{m}\rangle_{m}\neq0\;,\label{last}\end{eqnarray}
 which can be satisfied only if ${\langle \boldsymbol{N}}|e_{m}\rangle\neq0$ for
all eigenvectors $|e_{m}\rangle$ of the single chain Hamiltonian
$H$. QED.

While we have proved here that for equal time intervals the probability of success is converging to unity, in practice one may use
{\em optimal} measuring time intervals for a faster transfer~\cite{BURGARTH}. 
We also point out that timing errors may delay the transfer, but will not decrease
the asymptotic fidelity.

\subsection{Quantum chains with nearest-neighbors interactions}

It is worth noticing that Eq. (\ref{last}) is a very week condition,
which is satisfied for \emph{any} open nearest-neighbor quantum chain
as long as the transition amplitude $f_{1,N}(t)$ from Alice to Bob
(cf. Eq. (\ref{OUTPUT})) is not identical to zero. Let us prove
this by contradiction: assume there exists a normalized eigenvector
$\left|e_{m}\right\rangle $ of the single chain Hamiltonian $H$
such that \begin{equation}
{\langle  \boldsymbol{N}}|e_{m}\rangle=0.\end{equation}
Because $\left|e_{m}\right\rangle $ is an eigenstate, we can conclude
that also \begin{equation}
\left\langle e_{m}\left|H\right| \boldsymbol{N}\right\rangle =0.\label{eq:h_null}\end{equation}
If we act with the Hamiltonian on the ket in Eq. (\ref{eq:h_null})
we may get some term proportional to ${\langle e_{m}}| \boldsymbol{N}\rangle$ (corresponding
to an Ising-like interaction) and some part proportional to ${\langle e_{m}}|\boldsymbol{N-1}\rangle$
(corresponding to a hopping term; if this term did not exist, then
clearly $f_{1,N}(t)=0$ for all times). We can thus conclude that
\begin{equation}
{\langle e_{m}}|\boldsymbol{N-1}\rangle=0.\label{eq:steptwo}\end{equation}
Note that for a closed chain, e.g. a ring, this need not be the case,
because then also a term proportional to 
${\langle e_{m}}|\boldsymbol{N+1}\rangle={\langle e_{m}}|\boldsymbol{1}\rangle$
would occur. If we insert the Hamiltonian into Eq. (\ref{eq:steptwo})
again, we can use the same reasoning to see that \begin{equation}
{\langle e_{m}}|\boldsymbol{N-2}\rangle=\cdots={\langle e_{m}}|\boldsymbol{1}\rangle=0\end{equation}
 and hence $\left|e_{m}\right\rangle =0,$ which is a contradiction
to $\left|e_{m}\right\rangle $ being normalized. We thus conclude
that any nearest-neighbor Hamiltonian that can transfer quantum information
with nonzero fidelity (including the Heisenberg chains analyzed in
Refs.~\cite{BOSE,GIOVA}) is capable of efficient and perfect transfer
when used in the context of parallel chains.

\acknowledgments

VG acknowledges the support of the European Community under contracts
IST-SQUIBIT, IST-SQUBIT2, and RTN-Nanoscale Dynamics. DB acknowledges
the support of the UK Engineering and Physical Sciences Research Council through the 
grant GR/S627961/01 and the QIPIRC.


\begin{thebibliography}{10}
\bibitem{BOSE}S. Bose, Phys. Rev. Lett. \textbf{91}, 207901 (2003).
\bibitem{LLOYD}S. Lloyd, Phys. Rev. Lett. \textbf{90}, 167902 (2003).
\bibitem{GIOVA}V. Giovannetti and R. Fazio, Phys. Rev. A {\bf 71}, 032314 (2005).
\bibitem{key-38}V. Subrahmanyam, Phys. Rev. A \textbf{69}, 034304 (2004). 
\bibitem{PLENIO04}M.B. Plenio, J. Hartley, and J. Eisert, New J. Phys. \textbf{6},
36 (2004).
\bibitem{key-19}G. De Chiara, D. Rossini, S. Montangero, and R. Fazio, quant-ph/0502148.
\bibitem{key-11}M. Christandl, N. Datta, T. C. Dorlas, A. Ekert, A. Kay, and A. J.
Landahl, Phys. Rev. A \textbf{71}, 032312 (2005).
\bibitem{OSBORNE}T. J. Osborne and N. Linden, Phys. Rev. A \textbf{69}, 052315 (2004).
\bibitem{key-33}M. H. Yung and S. Bose, Phys. Rev. A \textbf{71}, 032310 (2005)
\bibitem{key-34}P. Karbach and J. Stolze, quant-ph/0501007.
\bibitem{key-17}M. B. Plenio and F. L. Semiao, New. J. Phys. \textbf{7}, 73 (2005).
\bibitem{key-12}Y. Li, T. Shi, B. Chen, Z. Song, and C. P. Sun, Phys. Rev. A \textbf{71}, 022301.
\bibitem{key-3}H. L. Haselgrove, eprint quant-ph/0404152.
\bibitem{BURGARTH}D. Burgarth and S. Bose, Phys. Rev. A \textbf{71}, 052315 (2005).
\bibitem{SHOR} C. H. Bennett and P. W. Shor, IEEE Trans. Inf. Theory
{\bf 44}, 2724 (1998).
\bibitem{PARALLEL}This is quite a common attitude in quantum information theory~\cite{SHOR} 
where successive uses of a memoryless channel are formally described by introducing
many parallel copies of the channel (see~\cite{GIOVA} for a discussion on the possibility
of applying this formal description to quantum chain models). Moreover for the case at hand
the assumption of Alice and Bob dealing with ``real'' parallel chain seems reasonable
also from a practical point of view~\cite{PARALLEL1}.
\bibitem{PARALLEL1}
N. Motoyama, H. Eisaki, and S. Uchida, Phys. Rev. Lett. {\bf 76}, 3212 (1996);
P. Gambardella, A. Dallmeyer, K. Maiti, M. C. Malagoli, W. Eberdardt, K. Kern, and C. Carbone, 
Nature {\bf 416}, 301 (2002).
\bibitem{key-32}D. Burgarth and S. Bose, New J. Phys. \textbf{7}, 135 (2005).
\bibitem{key-1}C. H. Bennett, D. P. DiVincenzo, and John A. Smolin, Phys. Rev. Lett
\textbf{78}, 3217 (1997).
\bibitem{INEQ}J. R. Schott, Matrix Analysis for Statistics, Wiley-Interscience (1996)
\bibitem{key-36}If $S$ is the similarity transformation that puts $T_{\ell}$ into the
Jordan canonical form, i.e. $J=S^{-1}T_{\ell}S,$ then $c$ is given
explicitly by $c=\| S\|\:\| S^{-1}\|$.







\end{thebibliography}
\end{document}